\documentclass[12pt,column,showpacs,pre,preprintnumbers,amsmath,amssymb,aps,standalone]{revtex4-2}
\usepackage{graphicx,dcolumn,bm,verbatim,subfigure,mathrsfs,csquotes,braket} 
\usepackage[usenames,
dvipsnames]{color}
\usepackage{epstopdf}

\begin{document}

\title{Spectral Symmetry Breaking of Electro-Acoustic Noise in Ferroelectric Materials}
\author{Dhiraj Sinha}
\affiliation{Department
	of Electrical Engineering  and Computer Science \\ Massachusetts Institute of Technology \\ Cambridge,
	MA, 02139 USA \\e-mail: dhiraj@mit.edu}

\date{\today}
\begin{abstract}
We present a novel analytical formulation on generation of electro-acoustic noise in ferroelectric materials where the thermal fluctuations induced polarization and acoustic modes distinctly contribute to noise. The conservative force fields associated with ferroelectric materials drive feedback of the spectral modes in the system. It results in symmetry breaking of the frequency spectrum of broadband noise leading to enhancement of specific modes which generate high amplitude narrowband noise. We describe the process under the theoretical framework of fluctuation-dissipation theorem in the context of ferroelectric materials.  We further note that such spectral feedback are absent in Johnson-Nyquist noise generation in conductors which have spectral symmetry.\\[12pt]
 \end{abstract}
\maketitle
\section{Introduction}
Our current understanding of noise in electronic systems is based on the initial work by Johnson and Nyquist which considers the Brownian motion of electrons under thermal fluctuation \cite{1,2} and has a uniform spectral density throughout the frequency band. A related kind of noise is shot noise which is associated with discrete nature of electrons \cite{3}. In an electronic circuit, noise is also generated by capacitors which is analyzed by equating the Boltzmann energy to capacitive energy storage \cite{4}. However, the nature of interaction of an electron with a capacitor which has a conservative force field is different from its interaction with a resistor which is dissipative in nature and its impact on the noise spectrum requires a more refined analysis. A related issue of vital importance is noise generated by ferroelectric elements which has widespread applications. Ferroelectric materials are used as an insulating layer in the Gate electrode of CMOS devices \cite{5}, as a filter in the front end part of wireless devices \cite{6} and as data storage elements \cite{7}. They are also used as actuator elements in scanning tunnelling microscope \cite{8,9}, atomic force microscope \cite{10,11}, disk drives \cite{12} besides having a widespread use as  sensors for infrastructure monitoring \cite{13,14}, magnetic field sensing \cite{15} and related applications where extremely high sensitivity is necessary \cite{16}. 

The objective of current work is to present a novel analytical model on noise generation under non-equilibrium thermal excitations in ferroelectric materials. The model incorporates the interaction between microscopic thermal fluctuations with acoustic and electromagnetic modes of a macroscopic system leading to a novel formulation of fluctuation dissipation theorem which considers the role of conservative force fields. An important goal of the current analysis is establishment of the generation of an asymmetric noise spectrum unlike Johnson-Nyquist noise, which has spectral symmetry.

Noise in ferroelectric materials has been explored in the context of non-equilibrium force fields. For example, Barkhausen noise in ferroelectric materials is generated when a ferroelectric sample is subjected to an electrostatic field inducing changes in the size of polarized domains in discrete, uneven steps  \cite{16a,16b,16c}. It is primarily centred around the harmonics of the driving field and shows geometric decay with a change in frequency. An additional source of noise in ferroelectric materials is an outcome of stress gradients along the domain wall under elastic interactions which can become dominant in switching in memory elements \cite{16d,16e,16f}. This noise has also been extensively studied in ferromagnetic materials where, under continuous change of external field, there is a sudden change in the size and orientation of magnetic domains which leads to changes in magnetization in discrete steps \cite{16g}. The microscopic clusters of magnetic spins becomes aligned to external field and increase their size under exchange interaction which leads to an avalanching effect. This eventually induces a magnetization jump and a sharp increase in noise \cite{16h}. Barkhausen noise has power law dependence similar to flicker noise as it decreases with an increase in frequency \cite{16i,16j}. 

An additional source of noise in dielectric materials is Telegraph noise or burst noise which has been studied in the context of transistors \cite{17a} and flash memories\cite{17b}. It appears as short bursts comprising of discrete voltage levels at random time intervals as a consequence of material defects or due to the trapping and discharge of electrons or ions along the thin film interfaces in a semiconductor crystal.  However, the exact physical origins of Telegraph noise is not well understood \cite{17c}.

In the current work, noise arising as a consequence of thermal fluctuations in the absence of an external driving force has been analysed which is aimed at supplementing the prior work on studies on noise in ferroelectric materials. It holds special significance in the context of ferroelectric components in electronic devices where thermal interactions play a dominant role. 
\section{Fluctuation induced Noise}
Spontaneous polarization below Curie temperature is a general feature of ferroelectric materials \cite{17}. Such polarization can also be induced under external electric fields  \cite{18}. The bound charges in the ferroelectric crystal are momentarily displaced under thermal fluctuations leading to charge separation and voltage generation. This results in capacitive energy storage which is eventually transformed into current and is momentarily stored as inductive energy before being dissipated as heat and getting injected into the heat bath. The thermal dissipation gives way to thermal fluctuation and the process is repeated.
The Van Dyke model for a ferroelectric material considers it as comprising of capacitor, resistor and inductor element (Fig.1a). Hence, the ferroelectric material interacting with a heat bath can be considered to be an inductor-capacitor-resistor circuit element where the voltage induced under thermal fluctuation is transformed into current and its dissipation into a heat bath which transfers the energy back to the system as thermal fluctuations (Fig.1b). The total energy of the ferroelectric material can be equated to the Boltzmann energy, $k_BT$ where $k_B$ is Boltzmann constant and $T$ is temperature. According to equipartition theorem, $k_BT=CV^2/2+LI^2/2$, where $C$ is Capacitance, $V$ is induced voltage, $I$ is current, $L$ is inductance, and the maximum value of noise voltage can be expressed as $V(t)=\sqrt{2KT/C}$.
\begin{figure}[h!]
	\centering \subfigure[]{
		\label{fig1a}
		\includegraphics[width=.6\linewidth]{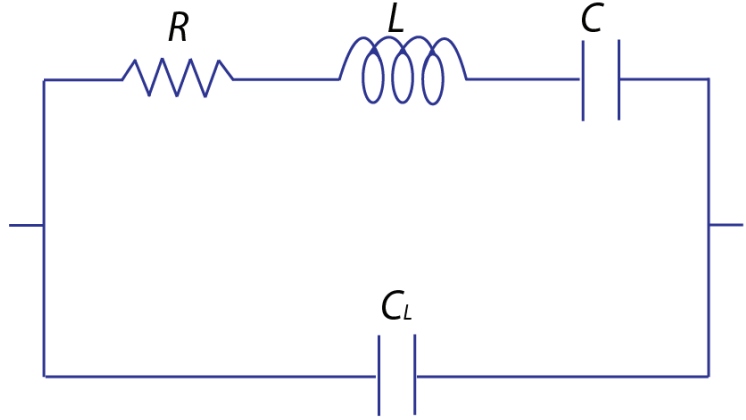}
	} \subfigure[]{
		\label{fig1b}
		\includegraphics[width=.7\linewidth]{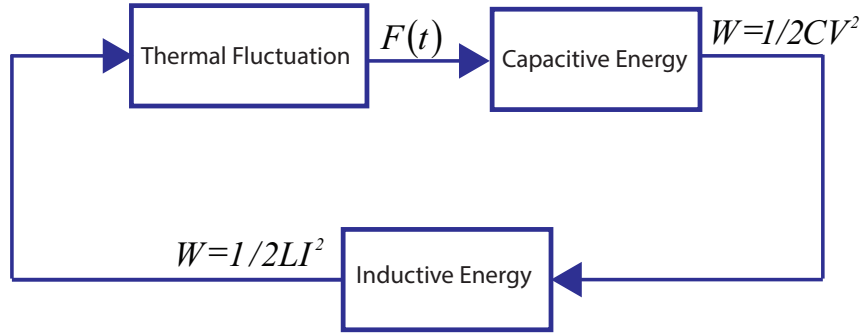}
	}
	\label{fig1}
	\caption{\small Ferroelectric material and its interaction with heat bath (a) Van Dyke Model of a piezoelectric crystal having lumped resistive ( $ R $), inductive ($ L $ ) and capacitive element ($ C $) along with lead capacitance $ C_L $. (b) The resistive element acts as a source of dissipative heat bath where energy is dissipated leading to generation of thermal fluctuations which leads to charge separation and voltage generation in a ferroelectric material. The capacitive energy generated is transformed into current leading to an increase in inductive energy which is soon dissipated as heat into the heat bath. The injected heat drives the thermal fluctuation.}
\end{figure}
Polarization modes induced in a ferroelectric crystal also leads to acoustic modes as a piezoelectric crystal lacks a centre of symmetry in its molecular structure \cite{18}. As the direction of force is reversed, the polarisation changes its direction which is expressed by the piezoelectric constitutive equation \cite{18},
$S_\mu=M_{\mu \nu}^ET_{\nu}+d_{\mu \nu}E_{\nu} $ where, $S$ is the strain, $M$ is mechanical compliance at constant electric field $E$, $T$ is stress, $d$ is piezoelectric coefficient, $\mu$ and $\nu$ are indices which run from $ 1 $ to $ 3 $ corresponding to spatial coordinates. In the simplest case where an electric field induces a displacement, the equation can be expressed as $S_\mu=d_{\mu \nu}E_\nu$. A piezoelectric crystal subjected to an electric field $E$ and mechanical stress $X$ develops an electrical polarisation $P$ and a strain $S$ which satisfies the relation,  
$(\partial P/ \partial X)_E=(\partial S/\partial E)_S$. The first term is the strain coefficient or the piezoelectric coefficient with the units m/V. Strain induced in the material under thermal fluctuations can result in induction of a finite amount of voltage.
The force on the charge centre as a consequence of thermal fluctuation is a sum of drag force, which takes it towards initial equilibrium and a fluctuating force with a net value of zero. This is described using Langevin equation \cite{19}, 
\begin{align} \label{e1}
m\frac{d^2\bm x}{dt}=-\beta  \frac{d \bm x}{dt}-\omega_m^2\bm x+\bm \Re(t)
\end{align}  
where, $m$ is the particle mass, $\bm x$ is its displacement, $ t $ is time, $\beta$  is the drag coefficient, $\omega_m$  is resonant frequency of the particle associated with conservative force field and $\bm \Re(t)$  is the random fluctuating function representing interaction with the heat bath. The net impact of the external force is displacement of atoms around the mean position which also leads to induction of charge. This can be visualised as an impulse function generated by a heat bath which convolutes with the impulse response of similar physical nature of piezoelectric crystal resulting in a noise spectrum. Taking the Laplace transform of Eq.~\ref{e1} and separating the terms \cite{20}, we can get the expression for displacement away from its mean position as,   
\begin{align} \label{e2}
x(s)=\frac{\Re(s)}{m(\omega_m^2+2\zeta_m \omega_ms+s^2)}  
\end{align}       
 Here, $ s=\omega \sqrt{-1} $, where $ \omega $ is the angular frequency of fluctuating field arising out of thermal interactions and $ \zeta_m=\beta/(2m \omega_m) $ is the mechanical damping coefficient. The random fluctuating function representing interaction with the heat bath has the property $\langle\Re(t_1)\Re(t_2)\rangle =2\beta k_BT \delta(t_1-t_2)$ where $t_1$ and $t_2$ are two instants of time. The spectral density can be written as its Fourier Transform,
\begin{align} \label{e3}
S_\Re(\omega)=\int_{-\infty}^\infty dt e^{j\omega t} \langle\Re(t_1)\Re(t_2)\rangle   
\end{align}
As the spectral density is proportional to $x|\omega|^2$,  we can write,   
\begin{align} \label{e4}
S_\Re(\omega)=\frac{2\beta k_BT}{m^2(\omega_m^2+2\zeta_m \omega_ms+s^2)^2}  
\end{align}     
Thus, the net displacement can be written as,
\begin{align} \label{e5}
x(\omega)=\frac{\sqrt{2\beta k_BT}}{m(\omega_m^2+2\zeta_m \omega_ms+s^2)}  
\end{align}   
Its time domain representaion is expressed as,
\begin{equation}\label{e6}
x(t)=\frac{ \sqrt{2\beta k_BT}}{\omega_m m\sqrt{(1-\zeta^2)}}e^{-\zeta\omega_m t}\sin{\omega_m\sqrt{(1-\zeta^2)} t}
\end{equation}
for $0\leq\zeta<1$. The expression in Eq. \ref{e6}  represents the net displacement of charge center under thermal fluctuation which contributes to noise. The term $\sqrt{2\beta k_BT }$ is a representative of Johnson-Nyquist noise, which is proportional to temperture as well as the drag coefficient which is a representative of the dissipation in the system. The analysis indicates the incorporation of the terms related to conservative force fields changes the physics of the system leading to addition of sinusoidal term which is dependent on resonant frequency. 

The displacement is expressed as, $x(t)=te^{-\omega t}$ for $\zeta=1$, and
\begin{equation}\label{e7}
x(t)=\frac{\sqrt{2\beta k_BT }}{2\omega_mm\sqrt{(\zeta^2-1)}}\Big[e^{-(\zeta-\sqrt{\zeta^2-1})\omega_m t}+e^{-(\zeta+\sqrt{\zeta^2-1})\omega_m t}\Big]
\end{equation}
for $\zeta>1$. 
The displacement of charge center can also be represented in terms of the Green's function for the given system such that, \begin{equation}\label{e8}
x(t)=\int_{-\infty}^{-\infty}dt'G(t,t')\frac{\sqrt{2\beta k_BT}}{m}
\end{equation}
where, $ G(t,t') $ is the Green's function expressed as,
\begin{equation}\label{e9}
G(t,t')=\int_{-\infty}^{-\infty}\frac{d \omega}{dt}\frac{e^{-i\omega(t-t')}}{\omega^2+2j\zeta\omega-\omega_m^2} 
\end{equation}
where $ j=\sqrt{-1} $. Thus, the displacement $ x(t) $ is the response of the sytem to thermal fluctuations which induces noise in the system.

By substituting, $\beta/m=2\zeta_m\omega_m$ in $\sqrt{2\beta k_BT}$  of Eq. \ref{e5}, we get $\sqrt{4\zeta_m\omega_mk_BT/m}$ where the Boltzmann energy per unit mass of a given molecule is denoted by the term $k_BT/m$. If the gram molecular mass of a molecule is $M_m$, then, $m=M_m/N_A$ where $N_A$ is the Avogadro constant. In a single gram of piezoelectric material, the total number of molecules is $N_A/M_m$ and the total Boltzmann energy in one gram of the molecule is the product of Boltzmann energy per unit mass ($k_BT/m$ or $k_BT/(M_m/N_A)$) and the number of molecules in $1$ g $(N_A/M_m)$ or $k_BT(N_A/M_m)^2$. Any change in displacement causes strain and we can define a voltage induced as $V(\omega)=\Im x(\omega)$, where $\Im$  is piezoelectric coefficient or strain voltage coefficient of the material which is of the order of $10$ pm/V for bulk zinc oxide \cite{21}. Hence, the total voltage induced in a sample of one gram of piezoelectric material is,
\begin{align} \label{e10}
V_m(s)=\frac{2\Im}{(\omega_m^2+2\zeta_m \omega_ms+s^2)}\sqrt{\zeta_m \omega_m k_BT}\left(\frac{N_A}{M_m}\right)  
\end{align}  
\begin{figure}[htbp]
	\subfigure[]{
		\label{Fig.2a}
		\includegraphics[width=.9\linewidth]{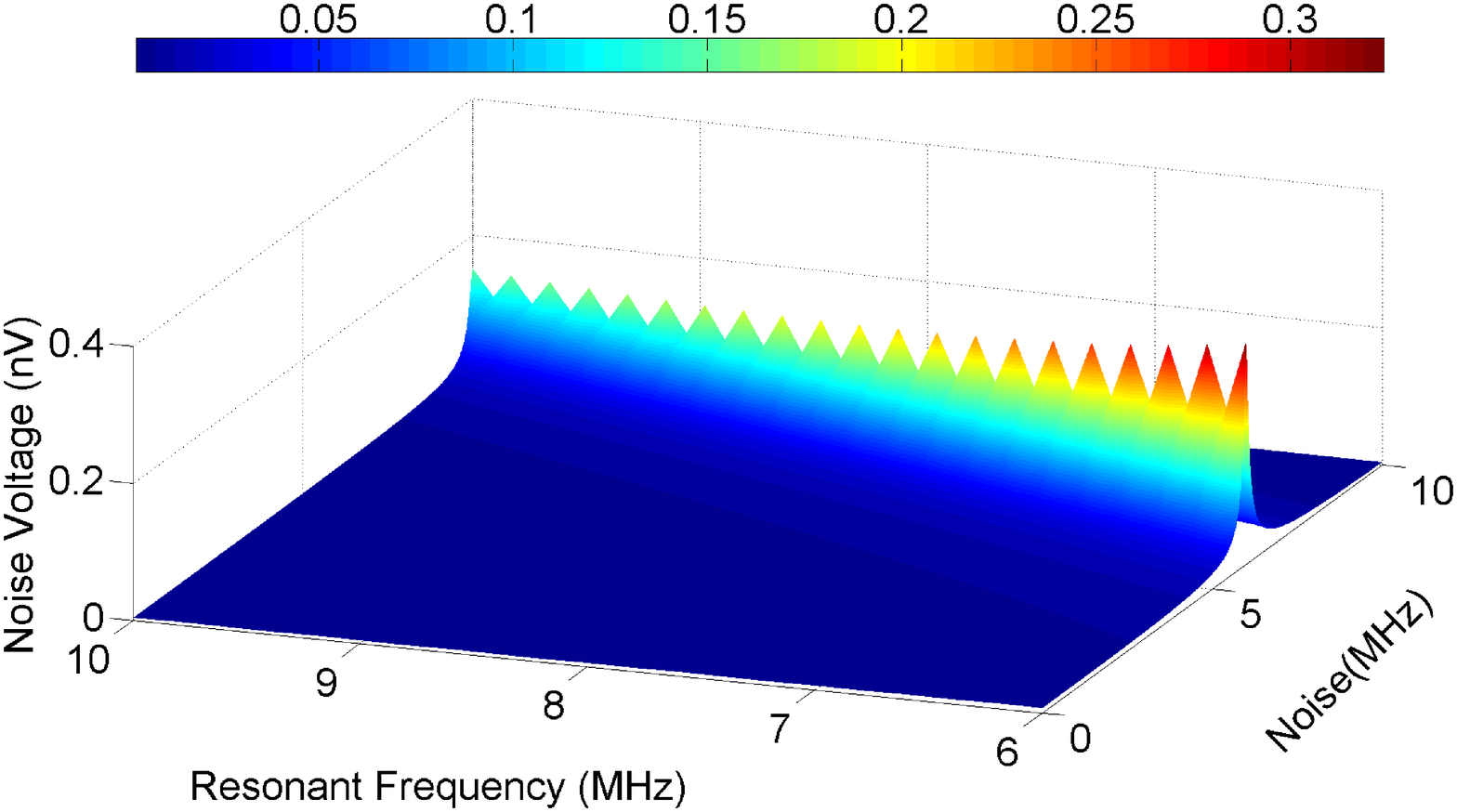}
	} \subfigure[]{
		\label{Fig.2b}
		\includegraphics[width=.9\linewidth]{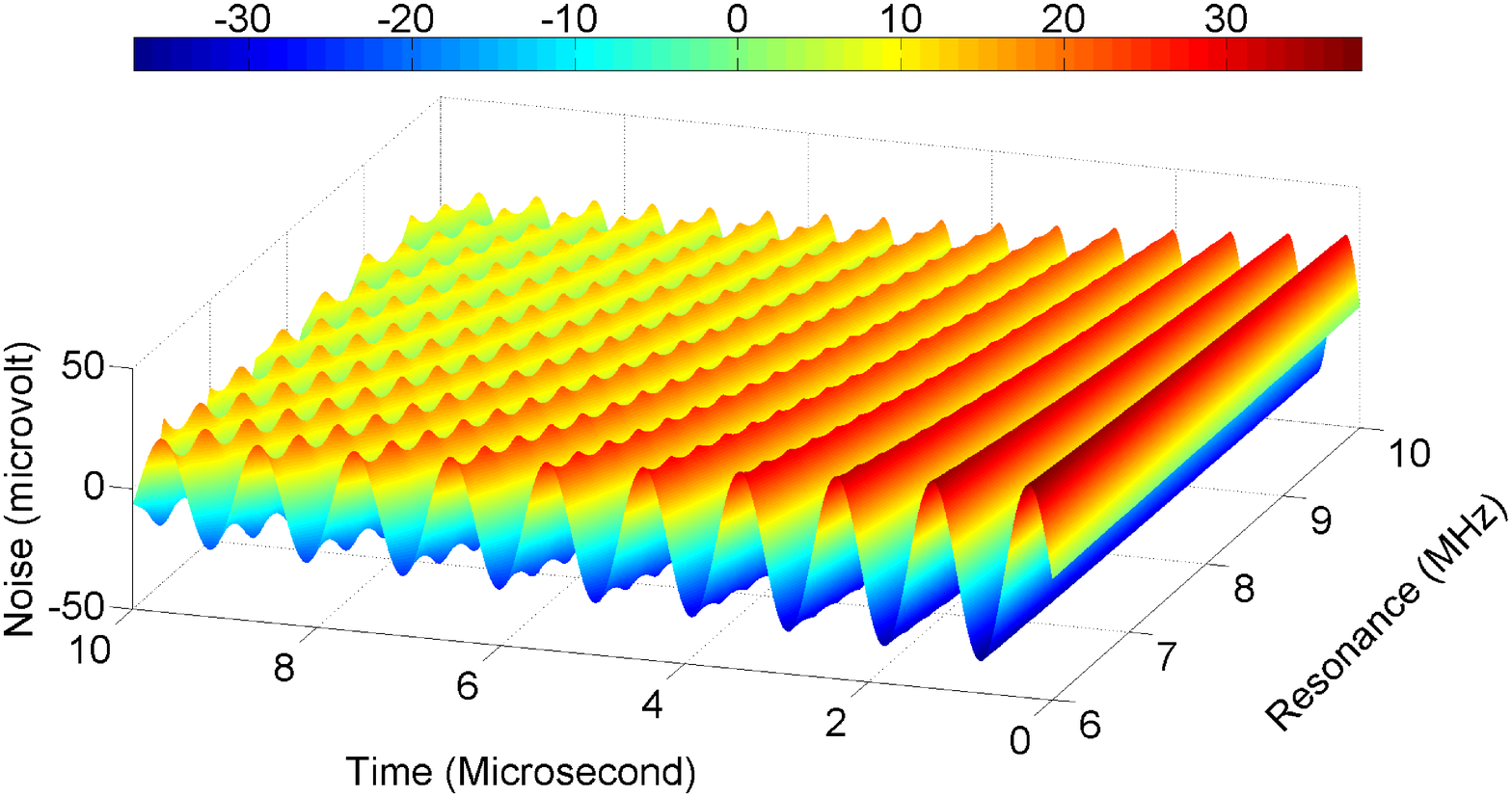}
	} 
	\label{fig2}
	\caption{\small Frequency and time domain representation of noise in a ferroelectric material. (a) The spectral distribution of electrical noise of a sample of ferroelectric material made of Zinc oxide of mass $ 1 $ g at room temperature as noise frequency is varied from $10$ kHz to $10$ MHz and resonant frequency is varied from $ 6 $ MHz to $ 10 $ MHz at a damping of 0.01. The colorbar indicates the magnitude of noise in nV. (b) The time domain representation of noise voltage over a resonance frequency variation from $6$ MHz to $10$ MHz as time is varied from $ 1 $ ns to $ 10 $ microsecond. Such high values of noise last only for short periods around the resonant modes and decrease exponentially with time. The colorbar indicates the magnitude of noise in $\mu$V.}
\end{figure}By considering the value of damping coefficient, $\zeta$=0.01, and the molecular mass of Zinc Oxide as 81.41 g \cite{16}, the voltage generated in frequency domain is graphically illustrated Fig. 2a. The noise frequency is assumed to vary from $10$ kHz to $10$ MHz and the resonant frequency of the piezoelectric material varies from 
 $ 6 $ MHz to $ 10 $ MHz with a damping coefficient of $0.01$. The noise voltage is in the range of fraction of a nanovolt which is below Johnson noise at a bandwidth of $1$ MHz. The Johnson noise voltage in a resistance of $R=1 $ $\Omega$, at a temperature of $T=300$ K, over a bandwidth of $\Delta f=1  $ MHz, is $ V=\sqrt{k_BT\Delta f R} =64.34$ nV. The relatively lower level of noise in ferroelectric materials in comparison to Johnson noise at a given temperature implies the latter's dominance in electronic circuits. However, since, the ferroelecltric noise increases with its mass and as the percentage of ferroelectric material increases, the overall noise figure achieve higher values.
  
In the time domain, the noise signal can be written as as the inverse Laplace Transform of Eq. \ref{e5}, 
\begin{align} \label{e11}
V_{out}(s)=\frac{2\Im}{\omega_m(1-\zeta^2)}
\sqrt{\zeta_m\omega_m k_BT}\left(\frac{N_A}{M_m}\right)e^{-\zeta\omega_m t}\sin{\sqrt{\omega_m(1-\zeta^2)} t}  
\end{align}  
Its graphical illustration is represented in Fig. 2b where the resonance frequency is assumed to vary from $6$ MHz to $10$ MHz and time is varied from $ 1 $ ns to $ 10 $ microsecond. The output noise signal voltage appears as bursts of sinusoidal signals with an amplitude in the range of $ 10 $ to $ 40 $ $\mu$V which exponentially decreasing with time. According to the Parseval's theorem, the net power of a signal in time and frequency domain remains the same, however, such high values of transient signals can affect the instantaneous sensor output leading to erroneous readings in sensitive measurements. 

In ferroelectric materials, the atoms vibrate at some specific resonant modes and the noise is mainly centred around the resonant modes. There is a significant level of attenuation of noise with an increase in frequency, hence the voltage and power spectrum has similar form factor. It appears that piezoelectric noise can be dominant at very short time scales at low frequencies but is attenuated at higher frequencies. The noise spectrum has spectral asymmetry and is different from the Fourier spectrum of Nyquist-Johnson noise, which is uniformly spread over a large frequency band. Thus, we can state that the spectral symmetry of noise is broken due to the presence of conservative force fields associated with the system.
\section{Feedback under Boundary Effect}
Electro-acoustic noise generated in a three dimensional block of piezoelectric material at a certain temperature has a certain number of resonant modes which undergo a change when finite boundary conditions are taken into account. The geometric shape has specific resonant modes corresponding to acoustic and electromagnetic boundary conditions. A heat bath interacting with the ferroelectric crystal injects fluctuating thermal field into the ferroelectric crystal. It leads to generation of acoustic waves displacing the charges leading to induction of time varying voltage (Fig. 3a). The voltage is transformed into current which is dissipated as heat into the heat bath. As the ferroelectric material has finite boundary conditions, a part of the current is fed back to it, which is transformed into acoustic and polarization modes. The impulse response of a given ferroelectric crystal can be considered to be interacting with a heat bath such that a part of its energy is fed back to the ferroelectric material while the remaining energy is fed to the heat bath which is the key source of fluctuating field (Fig. 3b).
\begin{figure}[htbp]
	\subfigure[]{
		\label{Fig.3a}
		\includegraphics[width=.8\linewidth]{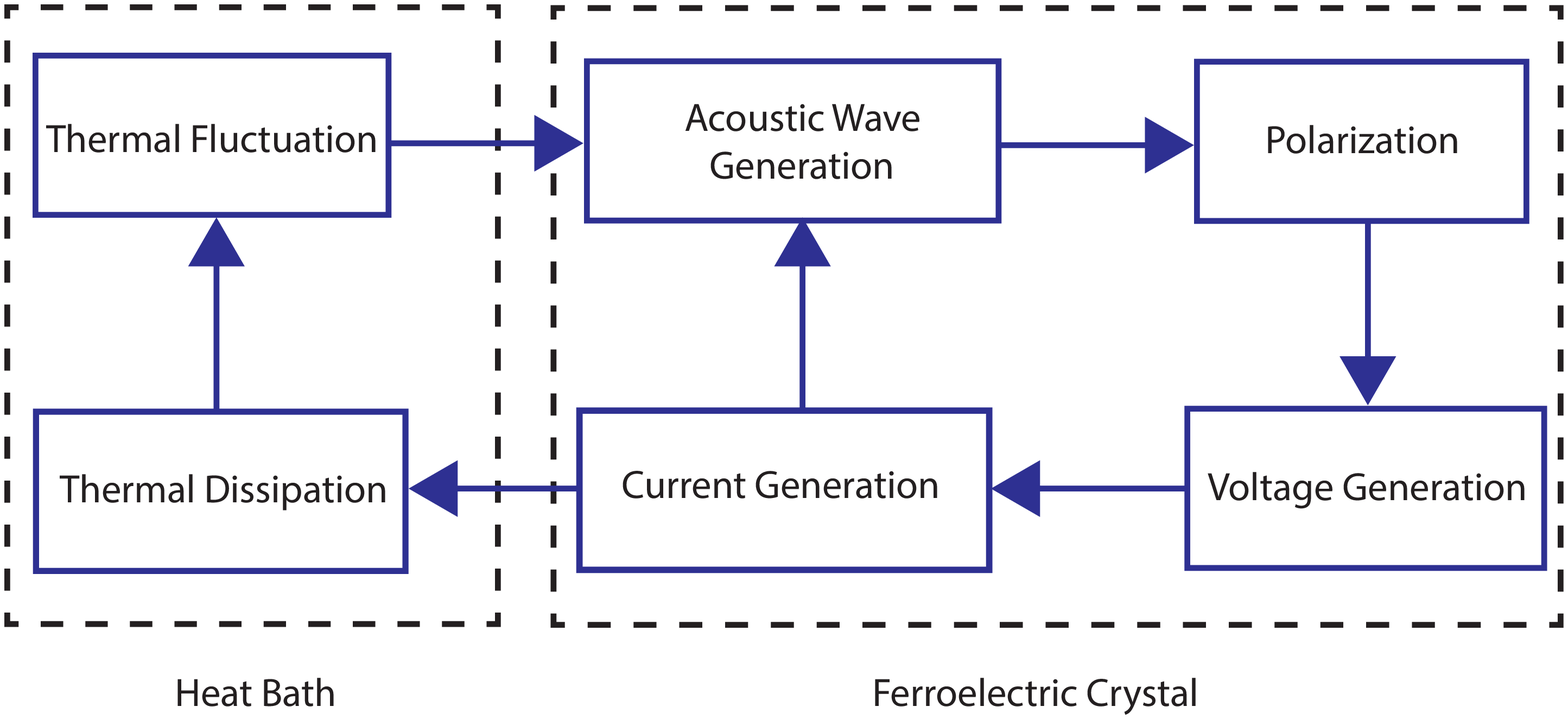}
	} \subfigure[]{
		\label{Fig.3b}
		\includegraphics[width=.8\linewidth]{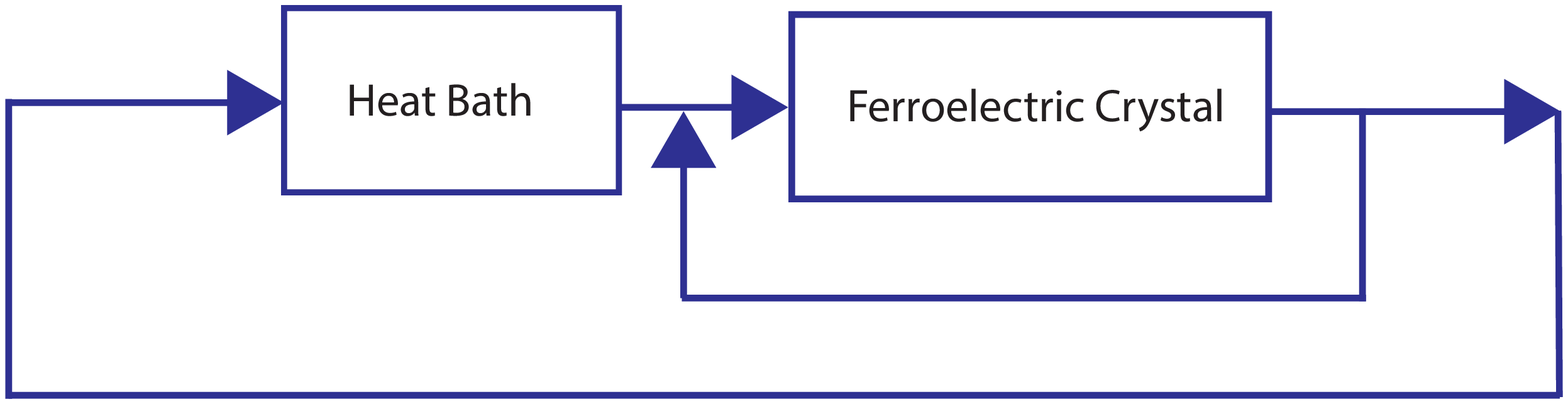}
	} 
\subfigure[]{
	\label{Fig.3c}
	\includegraphics[width=.8\linewidth]{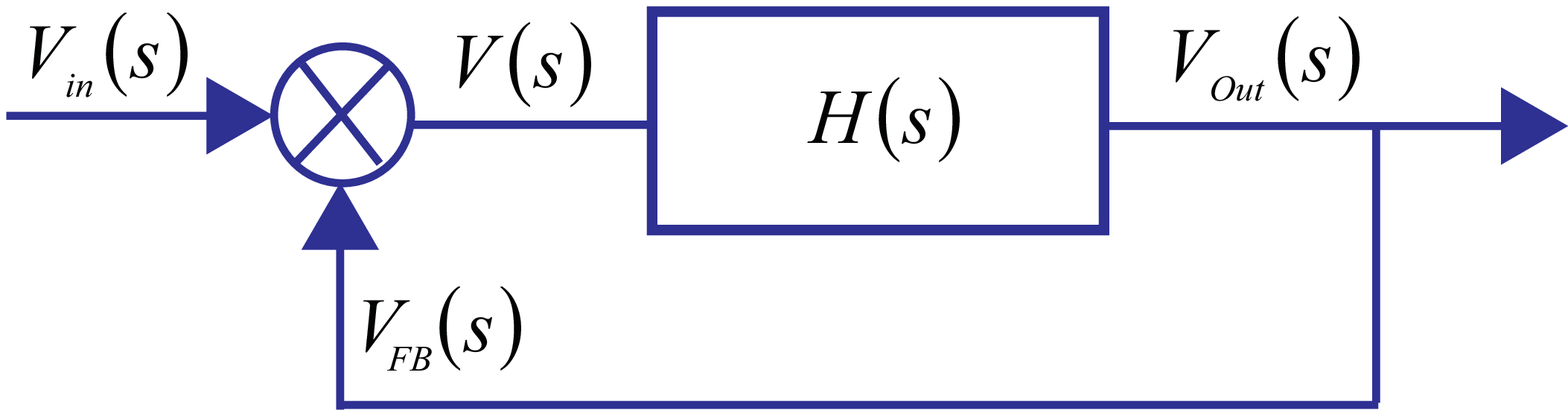}
} 
	\label{Fig. 3}
	\caption{\small Fluctuation dissipation relationship for a ferroelectric crystal. (a) Thermodynamic fluctuations applied to a ferroelectric crystal from a heat bath result in generation of acoustic wave which lead to polarization wave. The induced voltage is transformed into current which is eventually dissipated into the heat bath under feedback. (b) Thermodynamic fluctuations from a heat bath transfer energy to a ferroelectric crystal which is fed back to it and a part of the energy transferred to the heat bath which is the source of fluctuating thermal field. (c) The voltage $V(s)$ representing the spectral distribution of noise voltage is equal to $V_{in}(s)$ at time {t=0}. The output signal $V_{out}(s)$ is generated as a consequence of its interaction with the system's impulse response which is fed back to the input as $V_{FB}(s)$ where it is added to the noise input signal, $V_{in}(s)$. }
\end{figure}
The ferroelectric material along with acoustic and electromagnetic modes can be modelled as a multiple order system with a set of modes having a transfer function $ H(s) $ with a feedback path with a transfer function $ G(s) $ as shown in Fig. 3c. Here, the voltage $V(s)$ denotes the spectral distribution of noise voltage which equals
$V_{\text{\scriptsize in}}(s)$ at the start of the process. The output signal
$V_{\text{\scriptsize out}}(s)$ is generated as a consequence of its
interaction with the system's impulse response $ H(s) $ which is fed back to the
input as $V_{\text{\scriptsize FB}}(s)$ where it is added to the noise
input signal $V_{\text{\scriptsize in}}(s)$.
For simplicity, the ferroelectric material can be modelled as a second order system comprising a finite value of resistor $ R $, a capacitor $ C $ and an inductor $ L $ under the effect of an input voltage $ V_{in} $, which represents the impact of  acoustic-Brownian interactions inducing a time varying radio frequency field. The corresponding equation for the system can be written as~\cite{22},
\begin{equation}
\label{e12}
L\frac{dI(t)}{dt}+RI(t)+\frac{\int I(t)dt}{C}=V_{in}(t),
\end{equation}
where $L$ is the inductance, $R$ is resistance and $C$ is capacitance of
the system, $V_{in}$ is the input voltage (Fig. 4a). Substituting, $I(t)= CdV_{Out}(t)/dt$, in Eq. \ref{e12} and dividing both sides by $ LC $, we get, 
\begin{equation}
\label{e13}
\frac{d^2V_{Out}(t)}{dt^2}+2\zeta \omega_k\frac{dV_{Out}}{dt}(t)+\omega_k^2V_{Out}(t)=V_{in}(s)\omega_k^2
\end{equation}
Here, $\omega_k=1/\sqrt{LC_a}$, $\omega_k$ is resonant frequency of the system, $\zeta=R/(2L\omega_k)$ is its damping coefficient. It is graphically illustrated in Fig. 4b where it has been shown that $ V''_{Out}(s) $ and $ V'_{Out}(s) $ denote the second and the first derivative of output voltage. The output from the dissipative element, $ 2\zeta \omega V'_{Out} $ and the capacitor voltage $ V_{Out}(t) $ are fed back to the input.
\begin{figure}[h!]
	\centering \subfigure[]{
		\label{fig4a}
		\includegraphics[width=.8\linewidth]{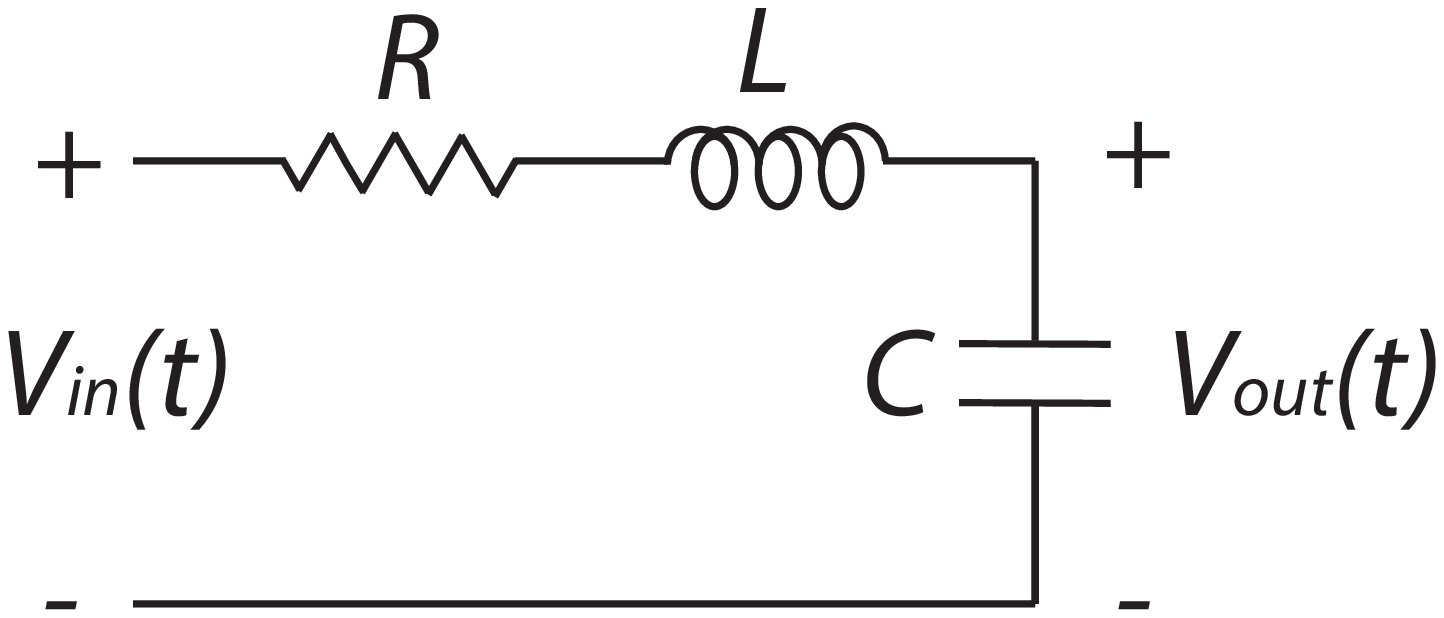}
	} \subfigure[]{
		\label{fig4b}
		\includegraphics[width=1\linewidth]{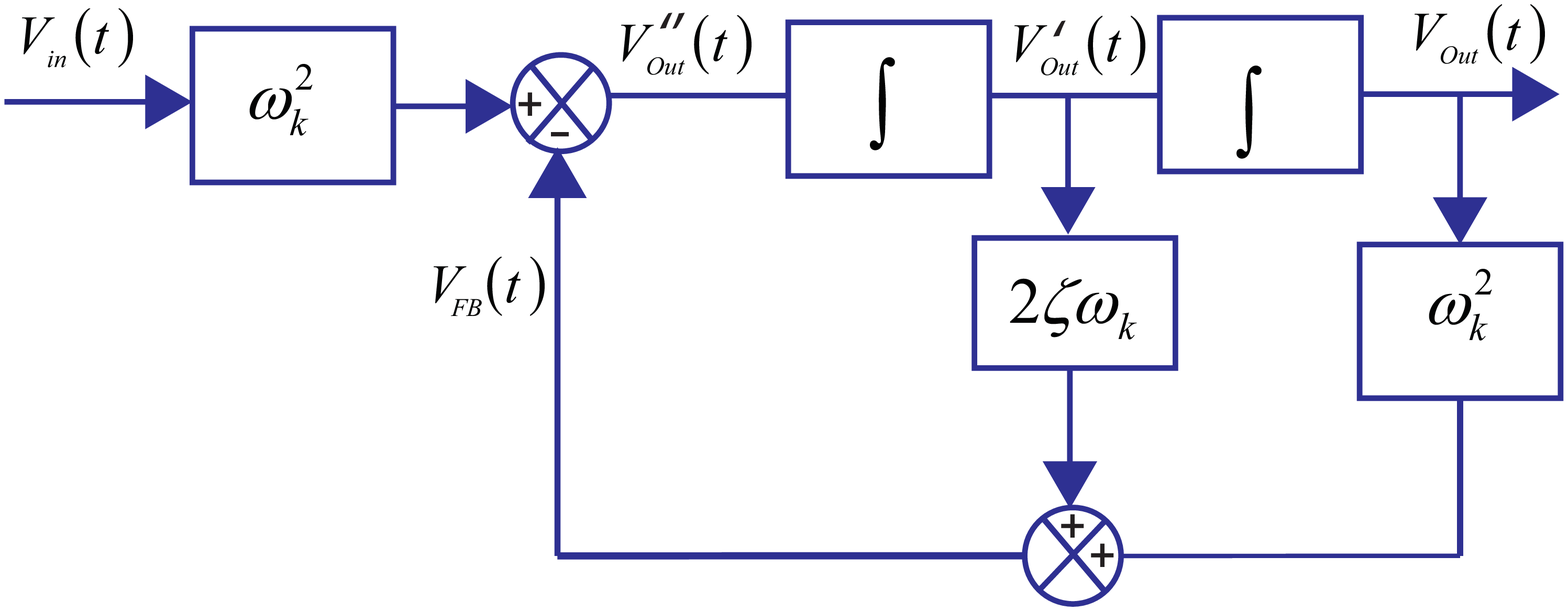}
	}  
	\caption{\small Circuit based model of feedback. (a) A resistor $ R $, capacitor, $ C $ and an inductor $ L $ in series represents the circuit equivalent model of a given specimen subjected to an input voltage radio frequency voltage $ V_{in} $. (b)  A block diagram based representation of Eq. \ref{e13} is shown here. $ V''_{Out}(s) $ and $ V'_{Out}(s) $ denote the second and the first derivative of output voltage. $ V_{Out} $ and $ V_{Out}(t) $ are fed back to the input.
	}
\end{figure}
Eq. \ref{e13} can be expressed using Laplace transform which leads to the expression expressed as, $ V_{in}(s)=I(s)(R+sL+1/Cs) $ where $ s=\omega\sqrt{-1} $ denotes the angular frequency of fluctuation associated with thermal interactions. The output voltage is $ V_{Out(t)}=(V_{in}/Cs)/(R+sL+1/Cs) $. We can write the output voltage drop as,
\begin{align} \label{e14}
V_{out}(s)=\frac{\omega_k^2}{s^2+2\zeta \omega_ks+\omega_k^2}V_{in}(s)  
\end{align} 
The resultant voltage expressed by Eq. \ref{e14}, expresses the response of
interaction of an input voltage generated as a consequence of thermal
fluctuations to the macroscopic parameters of the dielectric material defined
by its transfer function leading to an output voltage. It expresses the relationship between dissipation of radio frequency waves generated under acoustic fluctuations in ferroelectric material and can be considered as an expression of fluctuation dissipation theorem for a ferroelectric material. The process underscores the importance of selective mode enhancement under feedback.
As the acoustic and electrical modes co-exist in a ferroelectric material, the numerical values of $\omega_k$ are determined by the physical size and the corresponding electrodynamic boundary conditions.  The electrical resonant frequencies of a ferroelectric block considering it as a dielectric resonator are given by \cite{23}
\begin{align} \label{e15}
\omega_k=\frac{1}{a\sqrt{\mu \epsilon}}\sqrt{\left[\begin{array}{c}\chi_{np} \\ \chi'_{np}\end{array}\right]+\left[\frac{\pi a(2n+1)}{2d}\right]}
\end{align}
where $n$ is an integer, $d$ is its length, $a$ is radius, $\mu$ is permeability of the dielectric material, $\epsilon$ is its permittivity, $\chi_{np}$ and $\chi'_{np}$ are the parameters of Bessel function of the first kind which express the standing electromagnetic wave in the resonator in a cylindrical coordinate system \cite{24}. The propagation of mechanical waves displace the charge by finite amount leading to generation of voltage in the given material which keep getting reflected under the given set of boundary conditions. 

For a general system comprising of a number of resonant frequencies, $\omega_k$, where $k=$1,2,3….$N$, we can write\cite{23,24},
\begin{align} \label{e16}
V_{out}(s)=\prod_{i=1}^n\frac{V_{in}}{(s-\omega_k)}
\end{align}
For a second order system having an electrical resonant mode $\omega_k$ and damping  $\zeta$, the response can be written by equating $V_{in}(s)$ of Eq.~\ref{e14} with $V_m(s)$ of Eq. \ref{e6},                        
\begin{align} \label{e17}
V_{out}(s)=\frac{2\Im_{ij}\omega_m^{5/2}\sqrt{\zeta_m k_BT}(N_A/M_m)}{(\omega_m^2+2\zeta_m \omega_ms+s^2)(\omega_k^2+2\zeta \omega_ks+s^2)}
\end{align}
The noise voltage defined by Eq. \ref{e17}, expresses the fluctuation-dissipation theorem for a ferroelectric material where the electrical and mechanical modes are coupled. These modes play an integrated role in governing the overall noise along with the dissipative element. 

The output noise voltage is graphically illustrated in Fig. 5, where the electrical and mechanical damping coefficients are assumed to be equal, i.e. $ \zeta=\zeta_m=0.01 $. Initially, it is assumed that the system has a single mechanical resonant mode at $ 5 $ MHz while the system's mechanical resonant modes is assumed to lie between $6$ MHz to $10$ MHz in steps of $0.5$ MHz. The noise frequency spectrum is assumed to lie between $ 10 $ kHz to $ 10 $ MHz. Each of the resonant modes contribute to similar levels of noise as illustrated in Fig. 5a. The Illustration of noise voltage when the electrical resonant frequency is fixed at $5$ MHz and the electrical resonant frequency is varied from  $6$ MHz to $10$ MHz of nanovolt while the noise frequency spectrum lies between $ 10 $ kHz to $ 10 $ MHz is shown in Fig. 5b. The two sets of peaks correspond to each of these resonant frequencies and distinctly contribute to noise levels.

The dominance of selective modes in noise further indicates the broken symmetry of noise spectrum which has a physical nature different from Johnson-Nyquist noise. It offers a degree of freedom in terms of our ability to artificially control noise in ferroelectric system.	

\begin{figure}[htbp]
	\subfigure[]{
		\label{Fig.5a}
		\includegraphics[width=\linewidth]{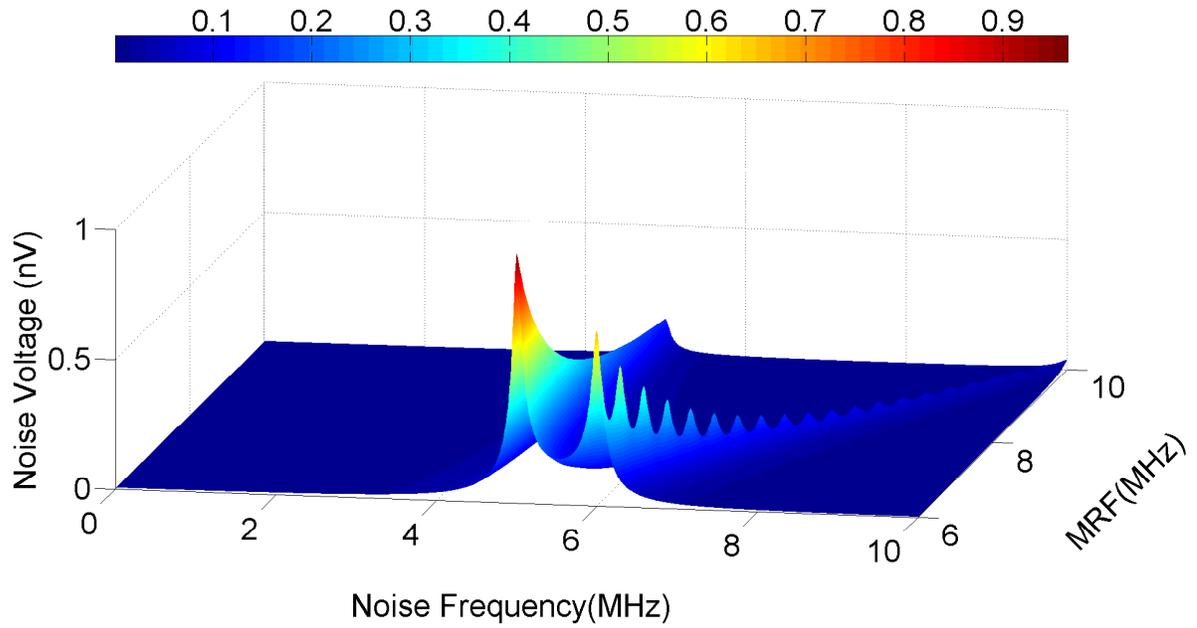}
	} \subfigure[]{
		\label{Fig.5b}
		\includegraphics[width=\linewidth]{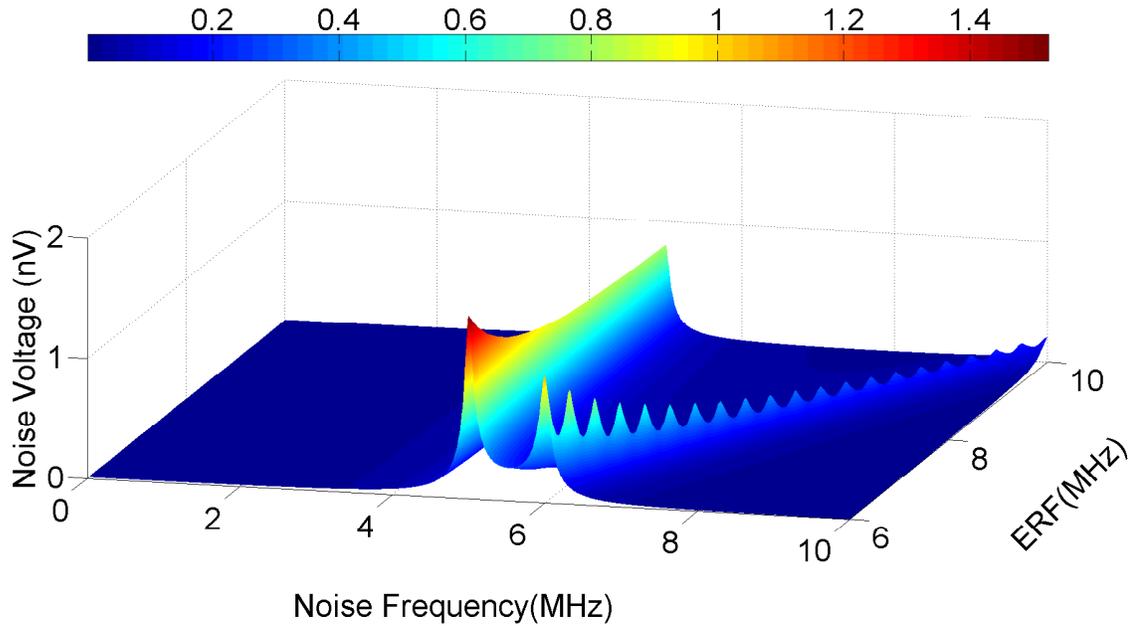}
	} 
	\label{Fig. 5}
	\caption{\small Noise in sample of ferroelectric material made of Zinc oxide of mass $ 1 $ g at room temperature (a) The noise frequency spectrum lies between $ 10 $ kHz to $ 10 $ MHz and the electrical resonant frequency (ERF) is fixed at $5$ MHz while the mechanical resonance frequency (MRF) varies from $6$ MHz to $10$ MHz. The two sets of peaks correspond to each of these resonant frequencies. The colorbar indicates the magnitude of noise in nV. (b) The mechanical resonant frequency is fixed at $5$ MHz and the electrical resonant frequency is varied from  $6$ MHz to $10$ MHz while the noise frequency spectrum lies between $ 10 $ kHz to $ 10 $ MHz. Each of the modes contribute to noise voltage. The colorbar indicates the magnitude of noise in nV.
	}
\end{figure}
An interesting effect results when the electrical and the mechanical resonant frequencies match. This is obtained by substituting, $\omega_k=\omega_m=\omega$ in Eq. \ref{e16}, we get,
\begin{align} \label{e18}
V_{out}(s)=\frac{2\Im_{ij}\omega_m^{5/2}\sqrt{\zeta k_BT}(N_A/M_m)}{(\omega_m^2+2\zeta \omega_m s+s^2)^2}
\end{align}
The noise voltage expressed by Eq. \ref{e18} is graphically illustrated in Fig. 6 which shows a drastic reduction in noise levels as the value of the denominator of Eq. \ref{e18} suddenly increases. Here, both the resonant modes are assumed to lie at $ 5 $ MHz while the noise spectrum lies between $ 1 $ kHz to $ 10  $ MHz. The reason behind sudden reduction of noise as the two resonant modes combine is the fact that energy is concentrated around a few resonant modes and remains screened from the surroundings. 
\begin{figure}[htbp]
	\subfigure[]{
		\label{Fig.6}
		\includegraphics[width=\linewidth]{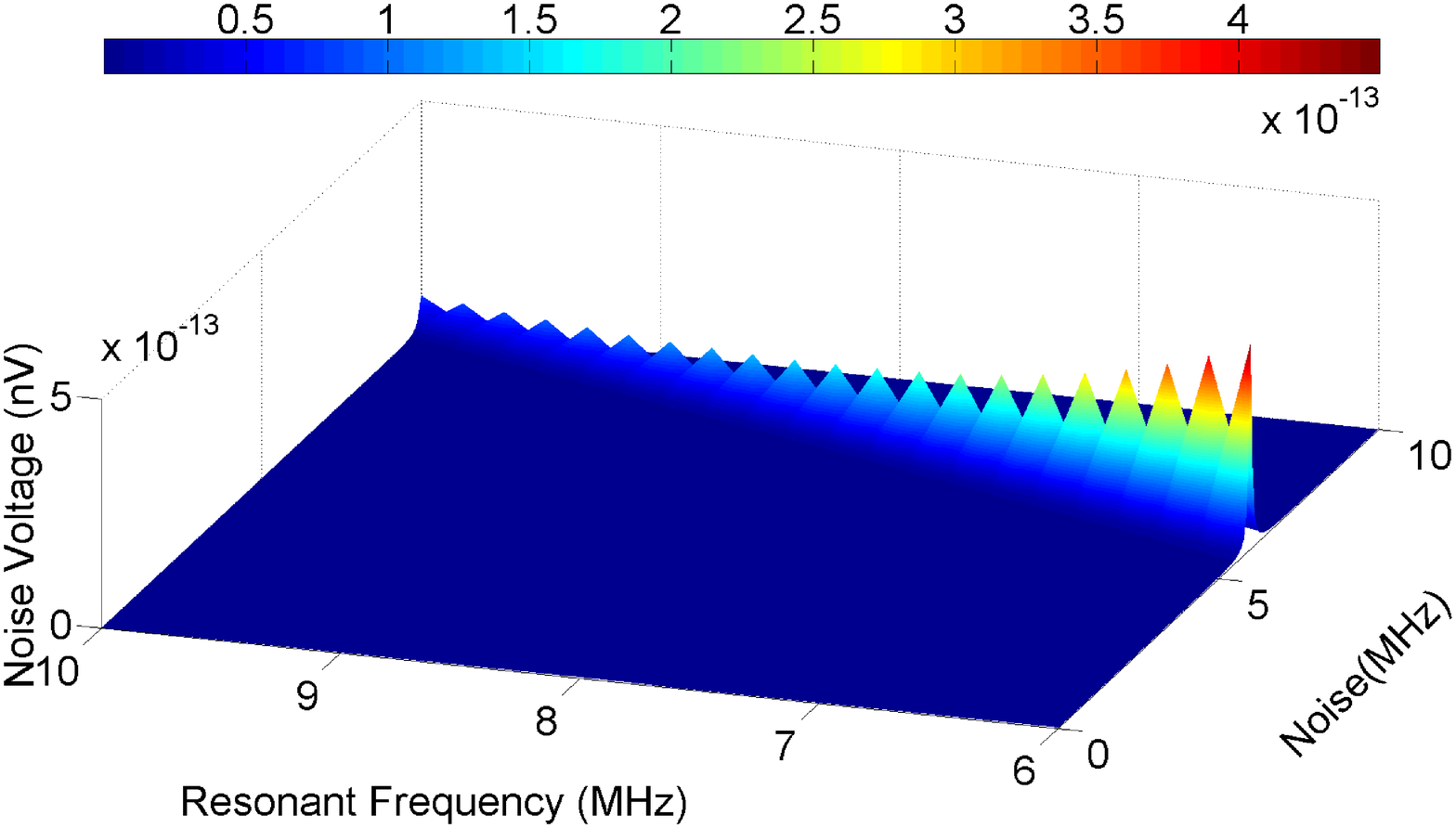}
	} 
	\label{Fig.6}
	\caption{\small Noise spectral density of a piezoelectric material. The spectral density shows a peak when the excitation frequency matches the resonant frequency at $ 5 $ MHz as the noise spectrum lies between $ 10 $ KHz to $ 10 $ MHz. 
	}
\end{figure}
The resultant voltage expressed by Eq.~\ref{e18} expresses the response of interaction of an input voltage generated as a consequence of thermal fluctuations to the macroscopic parameters of the ferroelectric material defined by its transfer function leading to an output voltage.  
\section{Discussion}
The fluctuation-dissipation theorem correlates stochastic fluctuations in a system to dissipation without making any specific reference to the type of material. It states that dissipation of energy by a system and generation of heat is associated with a reversible process which generates thermal fluctuations. The most notable example is Brownian motion where the kinetic energy of a particle is dissipated as a consequence of drag which induces fluctuations. It quantifies the fluctuations at thermal equilibrium to a system response under given set of perturbations. However, it was initially developed in the context of electrons in metals having resistance where thermal fluctuations of electrons result in fluctuation in voltage which is dissipated in the resistive element \cite{1,2}. Resistance dissipates electrical energy of a source leading to the generation of fluctuating voltage called the Nyquist Johnson noise. In the current case, the analysis has been extended to ferroelectric materials with some modifications which indicates that, unlike Johnson-Nyquist noise, which has a uniform spectral density; in ferroelectric materials, the output noise is determined by electrical as well as mechanical resonant frequencies of the system. In general, the noise decreases with an increase in frequency of external excitation. It can be stated that the spectral symmetry of noise frequencies is broken due to the enhancement of noise around the resonant frequencies as the the mechanical as well as electrical resonant frequencies add to distinct noise spectrum.

A key novelty of the current formulation is the fact that when the electrical and mechanical resonant frequencies of a ferroelectric material match, there is a sudden drop in noise level in terms of orders of magnitude as the noise generated by each of the degrees of freedom of the system is absorbed by the other. This observation holds  special importance for controlling noise in radio frequency filters which are made of ferroelectric materials comprising interdigital electrodes in the front end part of wireless devices and are designed to operate at frequencies between $20$ MHz to $3$ GHz \cite{25, 26}.
The frequency of resonant modes in a given material are defined by,
\begin{equation}\label{e19}
f=n\frac{v_s}{2L}
\end{equation}
where $ v_s $, is the velocity of acoustic wave, $L$ is the given length and $n$ is the harmonic number which has integral values. 
 In ferroelectric material like lithium niobate, the velocity of propagation of sound is in the range of $v_s=4000$ m/s \cite{27,28,29}.  If interdigital electrodes are developed on a film of lithium niobate, at a spacing of $ 4$ $ \mu $m, the mechanical resonant mode of the system would also be $ 500 $ MHz.  If the metallic interdigital electrodes possess a finger length of $3$ mm, the electrical resonance frequency would be $500$ MHz. Thus, a given device can be designed such that the electrical and mechanical resonances are matched which would lead to significant noise reduction. Such ferroelectric devices having closely spaced mechanical and electrical resonances are widely used in filters and sensors \cite{14} and in recent years have also found use in antennas \cite{30,31}. Ferroelectric material based devices are also widely used in microfluidic systems for cell separation \cite{32,33}. Thus, a better understanding of the mechanism of noise generation, offers additional degrees of freedom with regard to design of devices have a relatively higher level of sensitivity.
 
 Noise generation in a ferroelectric material under coupling between thermally induced acoustic modes and polarization modes with a heat bath, described in the current work, is focussed on the statistical outcome of interaction between electrons with the atoms of the solid crystal lattice. A more rigorous illustration of the phenomenon should consider electron-phonon interactions. Phonons, which are quantized representations of acoustic waves are comprised of two key components-acoustic modes, where the neighbouring atoms of the crystal are in phase and optical modes where, the neighbouring atoms are out of phase \cite{34}. In addition to these, there are zone boundary modes comprising of higher and lower energy modes and light modes. The current work, mainly incorporates the acoustic modes comprising of the long wavelength limit corresponding to the sound waves. A detailed study and analysis of electron-phonon interactions generated as a consequence of thermal fluctuations would incorporate the principles of quantum mechanics, while including the details of three dimensional lattice structure, and is beyond the scope of the current work.

 	A more rigorous analysis on electro-acoustic noise should incorporate charge fluctuations induced by Van der Waals interaction forces which are generated when the electron clouds of atoms interact at low interatomic distances. Such effects are dependent on 6th power of the interatomic distances and are classified as Keeson force, Debye force and London force depending on the nature of polarization of the participating molecules \cite{35}. However, in the current work, the key focus is on dependence of noise on macroscopic electrical  parameters of a ferroelectric material like capacitance and inductance which define the electrical resonant frequencies along with physical geometry and Young's modulus which control the mechanical resonant frequencies. These variables lead to an analytical model based on Transfer function formalism which effectively expresses a novel form of fluctuation-dissipation theorem for ferroelectric materials.
 
 The atomic and intermolecular interaction forces which play a role in determining the macroscopic electrical parameters have been neglected for simplicity of physical analysis. This is reasonable considering the fact that the dimensions of domain walls lie in the range of tens of nanometer \cite{36} to  a few hundred micrometers\cite{37} which is significantly smaller than the physics size of ferroelectric materials used in sensors which have macroscopic dimensions.

  An important objective of the current work is to lay down a theoretical framework which will motivate experimentalists to explore electro-acoustic noise at an empirical level in a manner independent of measurements on Nyquist Johnson noise, shot noise, Barkhausen noise and Telegraph noise. It incorporates physical variables like the dimensionality and boundary conditions which define the system's resonant frequency along with the role of capacitive and inductive elements which store energy.
  
The current formulation of the process of generation of electro-acoustic noise in ferroelectric materials has theoretical as well as empirical implications. The transfer function based approach can be used to calculate the response of the system under the impact of other physical effects which can contribute to noise. For example, the inverse Laplace Transform of Eq. \ref{e17} can be used to calculate the corresponding Green's function which can be instrumental in calculating the response of the system to other perturbations associated with thermal fluctuations which can contribute to noise.

In the context of ferroelectric based sensors and actuators for extreme environments like space applications, under-water acoustics, oil wells and chemical industries thermal effects are a critical determinant of device performance \cite{38, 39}. In the past, the process of breakdown in ferroelectric materials have been studied from the perspective of Maxwell-Boltzmann distribution of defects which produces dendritic microshorts \cite{40} or development of conductive tubular channels under electrical and mechanical loading \cite{41}. An analytical model on generation of noise, which could lead to high values of voltages and currents increasing the prospects of breakdown has not appeared in literature. The analytical model outlined in the current work can give critical variables towards taking suitable control measures which could ensure the continuous output from the sensor system. 
  \section{Conclusion}
We have shown that in ferroelectric materials, each of the mechanical as well as electrical resonant frequencies distinctly contribute to noise under interaction with the external heat bath. Thus, the fluctuation-dissipation theorem adopts a new form in ferroelectric materials where the conservative force fields drive feedback of energy during interaction with an external heat bath leading to spectral symmetry breaking and selective enhancement of specific modes. Such a feedback mechanism is absent in Nyquist Johnson noise in resistors which leads to spectral symmetry of noise. While it is well established that dissipative fields are mainly responsible for an increase in noise, the current work presents the role of conservative force fields as a source of noise. An additional finding is that by designing structures having similar electrical and mechanical resonant modes, the ferroelectric noise can be significantly reduced. The results of our analysis  open novel theoretical framework in determining the total noise in electronic devices and related systems where the density of piezoelectric components are increasing with miniaturisation.

Email:dhiraj@mit.edu

\end{document}